\documentclass{PoS}

\title{Calculation of the decay width of decuplet baryons}

\ShortTitle{Decuplet baryon resonances}

\author{Constantia Alexandrou$^{(a,b)}$, \speaker{Marcus Petschlies$^{(a,c)}$} \\ 
  $^{(a)}$Computation-based Science and Technology Research  
  Center, Cyprus Institute, 20 Kavafi Str., Nicosia 2121, Cyprus \\
$^{(b)}$Department of Physics, University of Cyprus, P.O. Box 20537, 1678 Nicosia, Cyprus 
  E-mail: \email{c.alexandrou@cyi.ac.cy}\\
$^{(c)}$Helmholtz-Institut fur Strahlen- und Kernphysik, Rheinische Friedrich-Wilhelms-Universit\"at Bonn, Nu{\ss}allee 14-16, D-53115 Bonn, Germany\\
  E-mail: \email{marcus.petschlies@hiskp.uni-bonn.de}
}
\author{John W. Negele, Andrew V. Pochinsky\\
  Center for Theoretical Physics, Laboratory for Nuclear Science and Department of Physics, Massachusetts Institute of Technology, Cambridge, Massachusetts 02139, U.S.A.\\
  E-mail: \email{negele@MIT.EDU}, \email{avp@mit.edu}
}
\author{Sergey S. Syritsyn\\
  RIKEN BNL Research Center, Brookhaven National Laboratory, Upton, NY 11973, USA\\
  E-mail: \email{ssyritsyn@lbl.gov}
}

\abstract{
  We calculate the coupling constant and decay width of the  decuplet to  octet baryon transitions in  lattice QCD using the transfer matrix method. The transition amplitude is related to the coupling constant and via the Fermi's Golden Rule to the decay width.
 The method is applicable for  near-degeneracy of the energy levels of initial and final states and, when this condition is fulfilled, yields 
a good estimate of the decay width. 
  We present  results using  a hybrid action  with domain wall valence quarks on a staggered sea  with $350\mev$ pion mass as well as for  a domain wall fermion action with $180\mev$ pion mass.
  We find $\Gamma\left( \Delta \to \pi\,N \right) = 119\,( 8)\,( 8)\mev$
  for the transition of Delta to pion-nucleon within the unitary domain wall setup. We also report values for the decay widths of the $\Sigma^*$ and $\Xi*$ baryons.
}

\FullConference{33rd International Symposium on Lattice Field Theory - LATTICE 2015\\
		July 14 - July 18, 2015\\
		Kobe, Japan}

\usepackage{amsmath}
\newcommand{\brackets}[1]{\langle #1 \rangle}

\newcommand{\sbrackets}[1]{\left[ #1 \right]}

\newcommand{\epow}[1]{\mathrm{e}^{#1}}

\newcommand{\mps}{m_{PS}}

\newcommand{\fermi}{\,\mathrm{fm}}

\newcommand{\mev}{\,\mathrm{MeV}}

\newcommand{\balign}{\begin{align}}
\newcommand{\ealign}{\end{align}}
\newcommand{\beq}{\begin{equation}}
\newcommand{\eeq}{\end{equation}}
\newcommand{\balignat}[1]{\begin{alignat}{#1}}
\newcommand{\ealignat}{\end{alignat}}

\newcommand{\bfig}{\begin{figure}}
\newcommand{\efig}{\end{figure}}

\newcommand{\bc}{\begin{center}}
\newcommand{\ec}{\end{center}}

\newcommand{\btab}{\begin{table}}
\newcommand{\etab}{\end{table}}

\newcommand{\bcom}{}

\newcommand{\bitem}{\begin{itemize}}
\newcommand{\eitem}{\end{itemize}}

\newcommand{\benum}{\begin{enumerate}}
\newcommand{\eenum}{\end{enumerate}}

\newcommand{\qvec}{\vec{q}}

\newcommand{\kvec}{\vec{k}}

\newcommand{\order}[1]{\mathcal{O}\left(#1\right)}

\newcommand{\refeq}[1]{(\ref{#1})}

\newcommand{\transferMatrix}{\mathcal{T}}
\newcommand{\chisqrPerDof}{\chi^2/\mathrm{dof}}

\begin{document}

\section{Introduction}

The calculation of baryon resonances from first principles in lattice QCD is a challenging task.
Physical decays cannot be observed directly in Euclidean space-time and
finite volume. One  method  devised to circumvent this no-go theorem in lattice QCD is given in  Ref.~\cite{Luscher:1985dn,Luscher:1986pf}, where
 a relation between energy shifts in finite volume and phase shifts in infinite volume was derived. The original method has been extended substantially
and applied successfully in studies of various meson-meson scattering processes.
A first application including baryons 
was presented in Ref.~\cite{Gockeler:2012yj}.
This finite volume method requires precise knowledge of energy levels in finite volume arising from either simulations with multiple lattice volumes or
calculations in multiple moving frames making this approach for baryons  
computationally intensive.

An alternative method has been proposed  in Ref. \cite{McNeile:2002az}, which is based on the computation of the  overlap of the decuplet and octet-pion  states on a finite lattice. This
 matrix element for the transition is then converted to the decay width of the resonance.
\begin{figure}[htpb]
  \centering
  \includegraphics[width=0.7\textwidth]{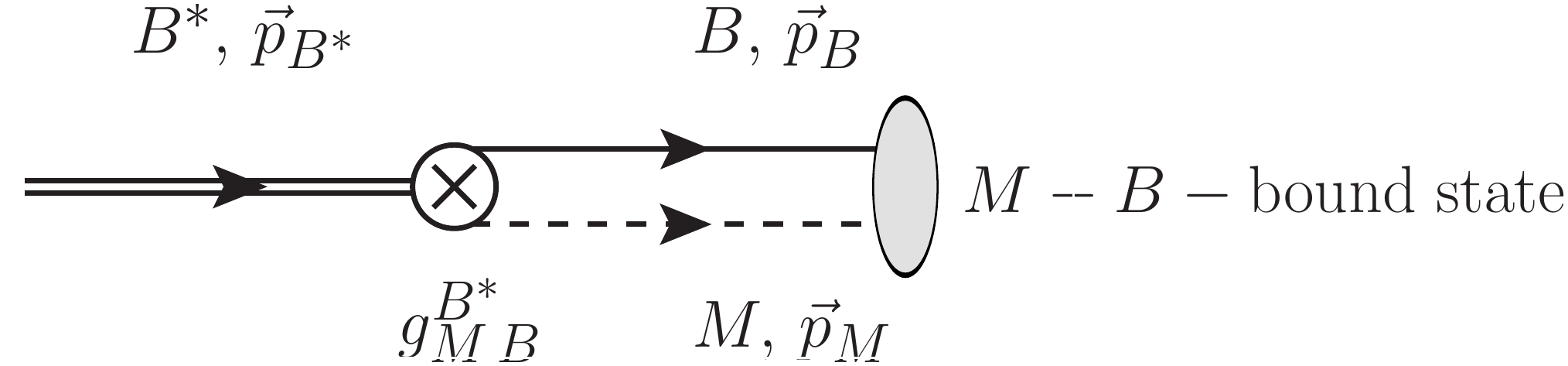}
  \caption{Diagrammatic representation for transition $B^* \to M\,B$ for a decuplet baryon $B^*$ to a meson - octet baryon $M\,B$ scattering state.
    The strength of the interaction is parametrized by the coupling $g^{B^*}_{M\,B}$.}
  \label{fig:Bstar_to_MB_diagram}
\end{figure}
The diagram in figure \ref{fig:Bstar_to_MB_diagram} depicts the approach considered here, namely a decuplet baryon resonance state $B^*$ couples
to a meson - octet baryon state $M\,B$. The strength of the interaction is parametrized by a coupling constant $g^{B^*}_{M\,B}$, which is
identified by matching to leading order effective field theory. In all the cases
considered in this work the meson $M$ is the pion.

\section{Transfer matrix method}

The transfer matrix method for a hadronic resonance state $B^*$ strongly decaying 
to a two-hadron scattering state $M\,B$ assumes that
 the energy levels of the initial and final states are firstly sufficiently close to each other and secondly sufficiently separated from the next higher lying state
in the spectrum. Up to exponentially suppressed corrections we can then consider the lattice transfer matrix in the sub-space spanned by 
$|\,B^*\rangle,\,|\,M\,B\rangle$,
\begin{eqnarray}
  \transferMatrix &=& \epow{-a\bar{E}}\,
  \begin{pmatrix}
    \exp^{-a\delta/2} & ax                & \cdots \\ 
      ax              & \epow{+a\delta/2} &        \\
      \vdots          &                   & \ddots \\
  \end{pmatrix}\,,
  \label{eq:transfer_matrix_subspace}
\end{eqnarray}
where $x$ denotes the desired transition amplitude. In this sub-space the lattice correlation function $\brackets{B^*,\,t_f\,|\,M\,B,\,t_i}$ can then be parametrized in terms
of the averaged energy of the states, $\bar{E} = \left( E_{B^*} + E_{MB} \right) / 2$, the energy gap $\delta = E_{MB} - E_{B^*}$ and the transfer matrix element
$x$ as in Eq.~\refeq{eq:lattice_overlap_parametrization}: using the notaton $T = t_f - t_i$ we thus write
\begin{eqnarray}
  \brackets{B^*,\,t_f\,|\,M\,B,\,t_i} &=& \brackets{B^*\,|\,\epow{-H\,T}\,|\,M\, B}
  = \brackets{B^*\,|\,\transferMatrix^{T/a}\,|\,M\,B} = ax\,\frac{\sinh(\delta\,T/2)}{\sinh(aT/2)}\,\epow{-\bar{E}\,T}\,,
  \label{eq:lattice_overlap_parametrization} \\
  &=& \sbrackets{ ax\,T + \order{\delta^2\,T^3} }\,\epow{-\bar{E}\,T} + \ldots\,.
  \label{eq:lattice_overlap_parametrization_linearized}
\end{eqnarray}
Eq.~\refeq{eq:lattice_overlap_parametrization} gives the sum of amplitudes for a single transition in Euclidean time space for $t_i \le t \le t_f$ for a time interval $T = t_f - t_i$.
For sufficiently small $T\,\delta$ the hyperbolic sine can be linearized and we have the leading time dependence as given in Eq.~\refeq{eq:lattice_overlap_parametrization_linearized} where
the ellipsis represent contributions from multiple transitions, excited states and mixing with other states.

\section{Lattice calculation}
We perform the calculation using two gauge field ensembles: one with a hybrid action using domain wall valence quarks on a sea of $N_f=2+1$ staggered fermions at $\mps = 350\mev$~\cite{Bernard:2001av}
and one with $N_f=2+1$   domain wall fermions at $\mps = 180\mev$ \cite{WalkerLoud:2008bp}. 
Initial results for the hybrid action were reported in Ref.~\cite{Alexandrou:2013ata} and recent results for the unitary domain wall fermion action have been given in Ref.~\cite{Alexandrou:2015hxa}.
\begin{table}
  \begin{center}
  \begin{tabular}{l|ccccc|cc}
    \hline
    action & $ L^3\times T$ & $\mps / \mev$ & $a / \fermi$ & $L / \fermi$ & $L_5/a$ & $N_\mathrm{conf}$ & $N_\mathrm{src}$ \\
    \hline
    hybrid  & $28^3\times 64$ & $350$ & $0.124$ & $3.4$ & 16 & 210 & 4 indep. \\
    unitary & $32^3\times 64$ & $180$ & $0.143$ & $4.5$ & 32 & 254 & 4 coh.   \\
    \hline
  \end{tabular}
\end{center}

  \caption{Parameters for the two ensembles used in this work. The last two columns give the number of gauge configurations analyzed
and the number of measurements per gauge configuration. For the hybrid action these consist of 4 independent measurements per configuration, whereas for unitary action 4 measurements are done
using the coherent sequential source approach.}
  \label{tab:ensemble_data}
\end{table}
We compile the parameters for the two ensembles used in our calculations in Table~\ref{tab:ensemble_data}. .

\begin{figure}[htpb]
  \centering
  \includegraphics[width=0.49\textwidth]{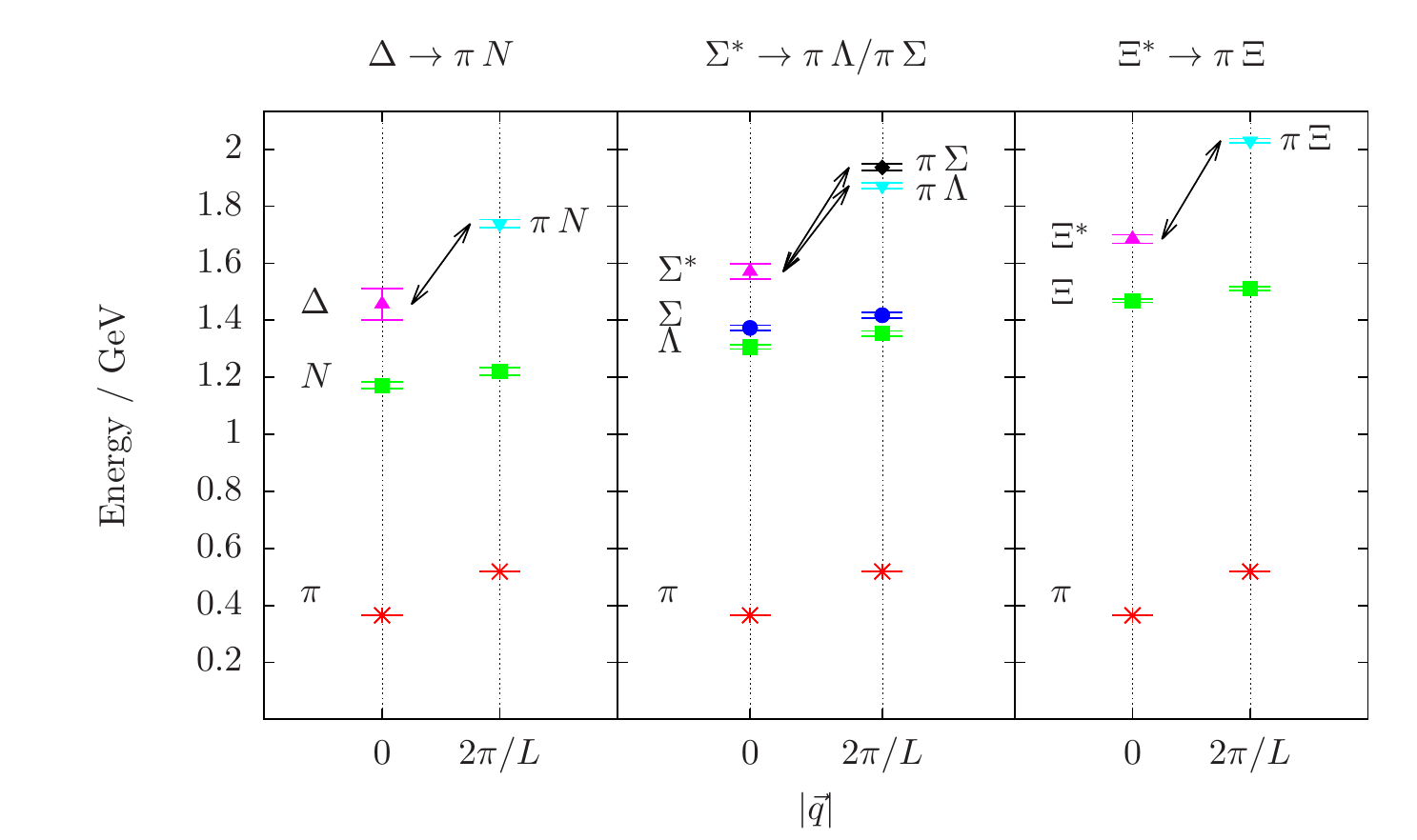}
  \includegraphics[width=0.49\textwidth]{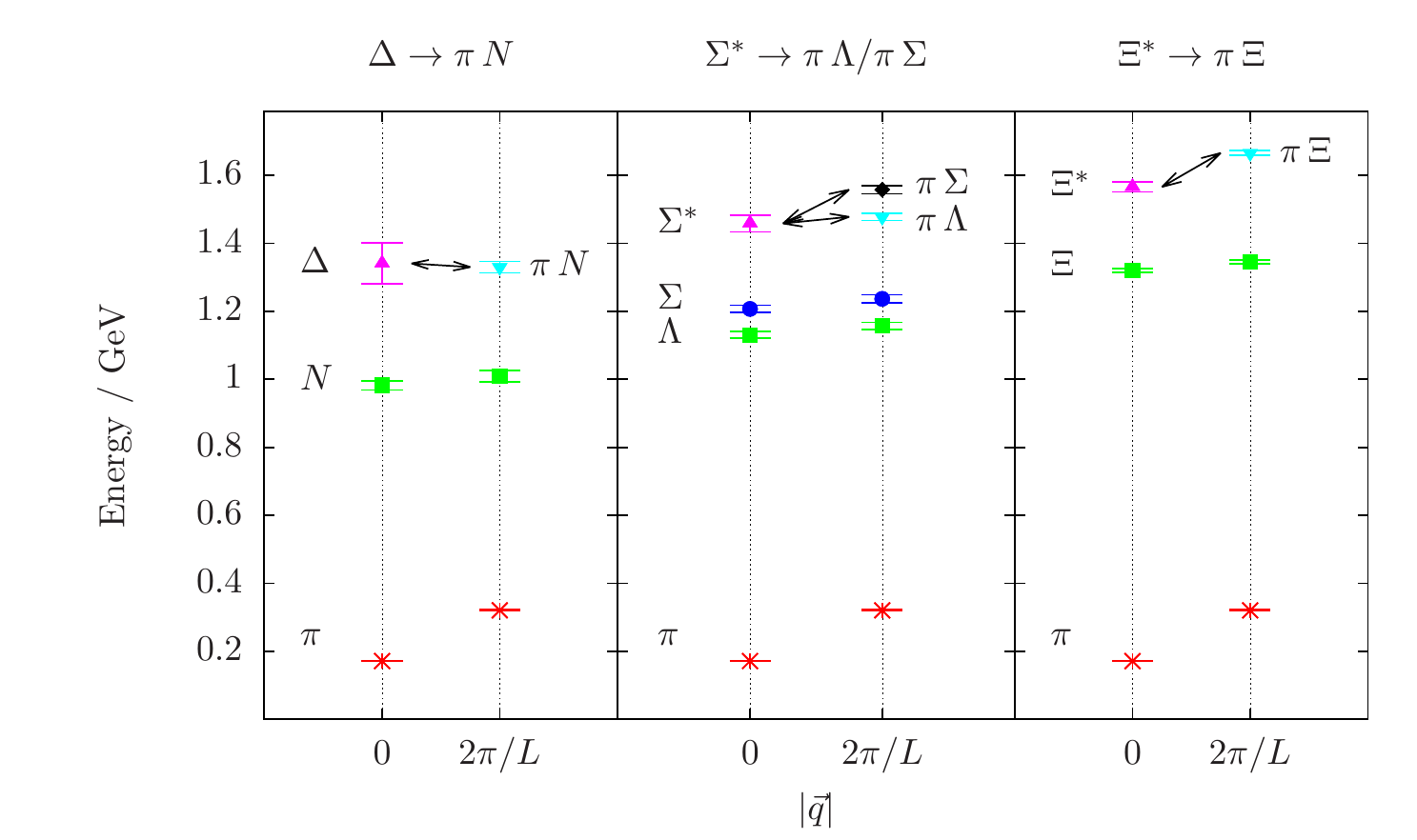}
  \caption{The low lying energy levels for the decuplet  $\Delta,\,\Sigma^*,\,\Xi^*$ and octet  $N,\,\Lambda,\,\Sigma^*,\,\Xi$ baryons and $\pi$ for the
  hybrid (left) and unitary (right) action. In each of the three panels, the left column displays the values of the  masses (zero spatial momentum), whereas the right column shows
  the energy level for one unit of momentum $|\qvec| = 2\pi/L$. The 2-hadron energy levels are approximated by the sum of the single-hadron energies $E_{M\,B} = E_{M} + E_{B}$. 
  The transitions given in Eq.~\protect\refeq{eq:decuplet_to_pion_octet_transitions} are marked by black arrows.}
  \label{fig:spectrum_light_strange}
\end{figure}

We  consider the following  transitions 
\begin{eqnarray}
  \Delta &\to& \pi\,N\,,\quad \Sigma^* \to \pi\,\Lambda\,,\quad \Sigma^* \to \pi\,\Sigma \,,\quad \Xi^* \to \pi \,\Xi
  \label{eq:decuplet_to_pion_octet_transitions}
\end{eqnarray}
in the positive parity channel. In all four cases the energy level of the initial and final states are the lowest lying states in the spectrum.
Fig.~\ref{fig:spectrum_light_strange} shows the relevant energy levels as measured in our lattice calculation 
As can be seen,  in the case of the hybrid action there is a significant energy gap between the $B^*$ and $M\,B$ states for all transitions. For the case of  the unitary action the energies are much closer
and thus the applicability criteria of the transfer matrix approach are better satisfied. This is particularly so for the case of $\Delta \to \pi\,N$ and to a lesser degree still for $\Sigma^* \to \pi\,\Lambda$,
where we have a near-degeneracy of energy levels. 

To extract the amplitude $x$ in Eqs.~\refeq{eq:lattice_overlap_parametrization} and \refeq{eq:lattice_overlap_parametrization_linearized} we construct a suitable ratio
of single- and two-hadron correlation functions
\begin{eqnarray}
  R^{B^*}_{M B,\,k}\left( t_f-t_i,\,\qvec,\,\kvec_f,\,\kvec_i\right) &= \frac{
    C_{B^*-M B}^k(t_f-t_i,\qvec)
  }{ 
    \sqrt{ C_{B^*-B^*}(t_f-t_i) \times C_{M B - M B}(t_f-t_i,\kvec_f,\kvec_i) }
  }\,.
  \label{eq:ratio_definition}
\end{eqnarray}
Here $C_{B^* - B^*}$ and $C_{MB - MB}$ are the 2-point correlators for the states $B^*$ and $MB$, respectively, and $C^k_{B^* - MB}$ is the transition 2-point correlation function.
The two-hadron interpolating field, taken as the product of single-hadron interpolating fields,
 is approximated by the quark-disconnected (direct) diagram, which amounts to $C_{MB - MB} \approx C_{M-M}\cdot C_{B-B}$.
We project to zero total momentum and assign one unit of relative momentum $|\qvec| = |\kvec| = 2\pi/L$ to the $M-B-$state. The explicit projection to  spin-3/2
and taking the meson to be the pion will then ensure the desired predominant overlap of the decuplet spin-3/2 state with the $M-B-$state at angular momentum $l = 1$ and thus positive parity.
The ratio is then averaged over all six  momentum directions as well as forward and backward propagation.

\begin{figure}[htpb]
  \centering
  \includegraphics[width=0.8\textwidth]{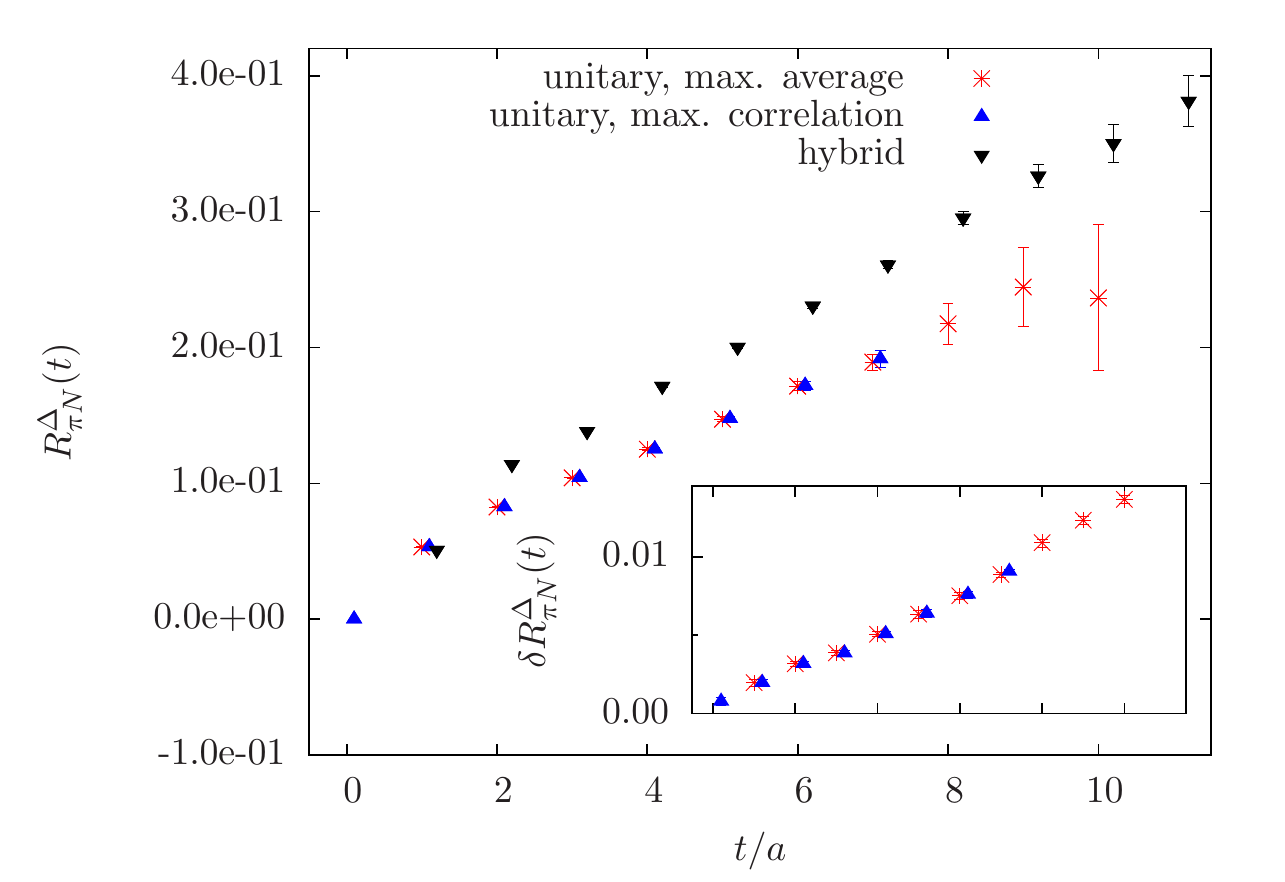}
  \caption{Ratio $R^{\Delta}_{\pi\,N}$ for the hybrid (black downward triangles) and unitary action. For the latter we show two different
  ways to combine data for individual momenta and forward and backward propagation; the data marked by red stars is used in the rest of the calculation.
  The detailed plot shows the time-dependence of the uncertainty of the ratio.
}
  \label{fig:ratio_deltapp2piN_comparison}
\end{figure}
Fig.~\ref{fig:ratio_deltapp2piN_comparison} shows our results for the ratio $R^{\Delta}_{\pi\,N}$ as an example. For both hybrid and unitary action we observe
a time interval, where the time-dependence of the ratio is consistent with the linearized form in Eq.~\refeq{eq:lattice_overlap_parametrization_linearized}.
We note, that the time-interval, in which a meaningful estimate of the ratio is available is limited by i) the oscillatory behavior of the domain wall fermion correlation functions at small times~\cite{Syritsyn:2007mp}, and ii) by the stochastic error of the correlator $C_{B^* - B^*}$ that grows exponentially with time and  which enters the ratio 
with an inverse square root.
To extract the overlap $x$ we fit the ratio using the fit ans\"atze
\begin{eqnarray}
  f_1(t) &=& c_0 + c_1\,\frac{\sinh\left(c_2\, t / 2a \right)}{\sinh\left( c_2/2 \right)} \,,\quad
  f_2(t) = c_0 + c_1\,\frac{t}{a} + c_2\,\left( \frac{t}{a} \right)^3 + \ldots
  \label{eq:fit_ansaetze}
\end{eqnarray}
with $ax = c_1$. $f_1(t)$ is derived from the expected time-dependence of the ratio when the lowest two states  with an energy gap dominate,
amended by a time-independent off-set, which accounts for contributions from excited states. $f_2(t)$ is an expansion of $f_1(t)$  assuming a small energy gap and  contains
free odd-power polynomial coefficients from which we keep up to the three leading ones, nemaly $c_0,\,c_1,\,c_2$. 

\begin{figure}[htpb]
  \centering
  \includegraphics[width=0.47\textwidth]{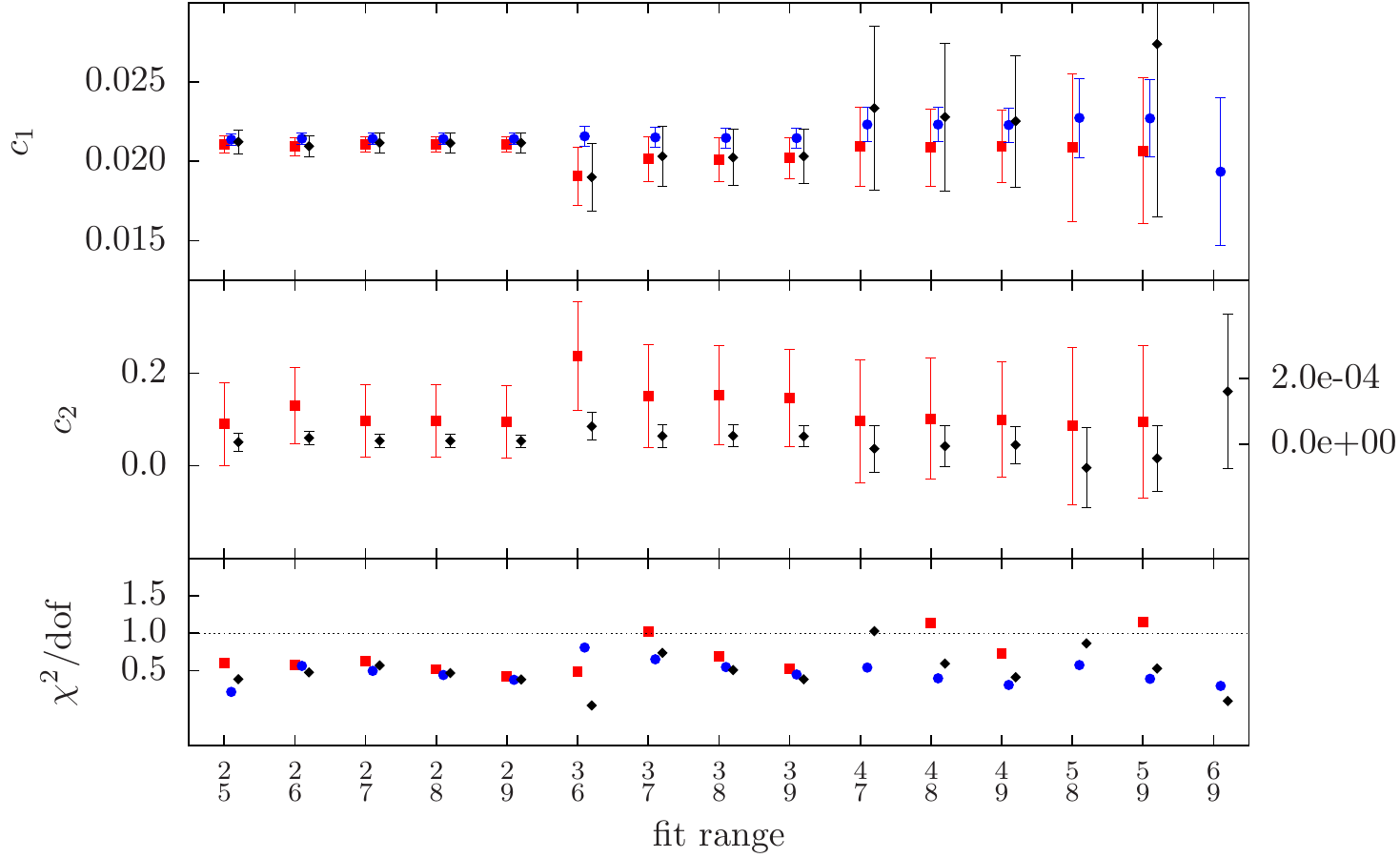}
  \includegraphics[width=0.47\textwidth]{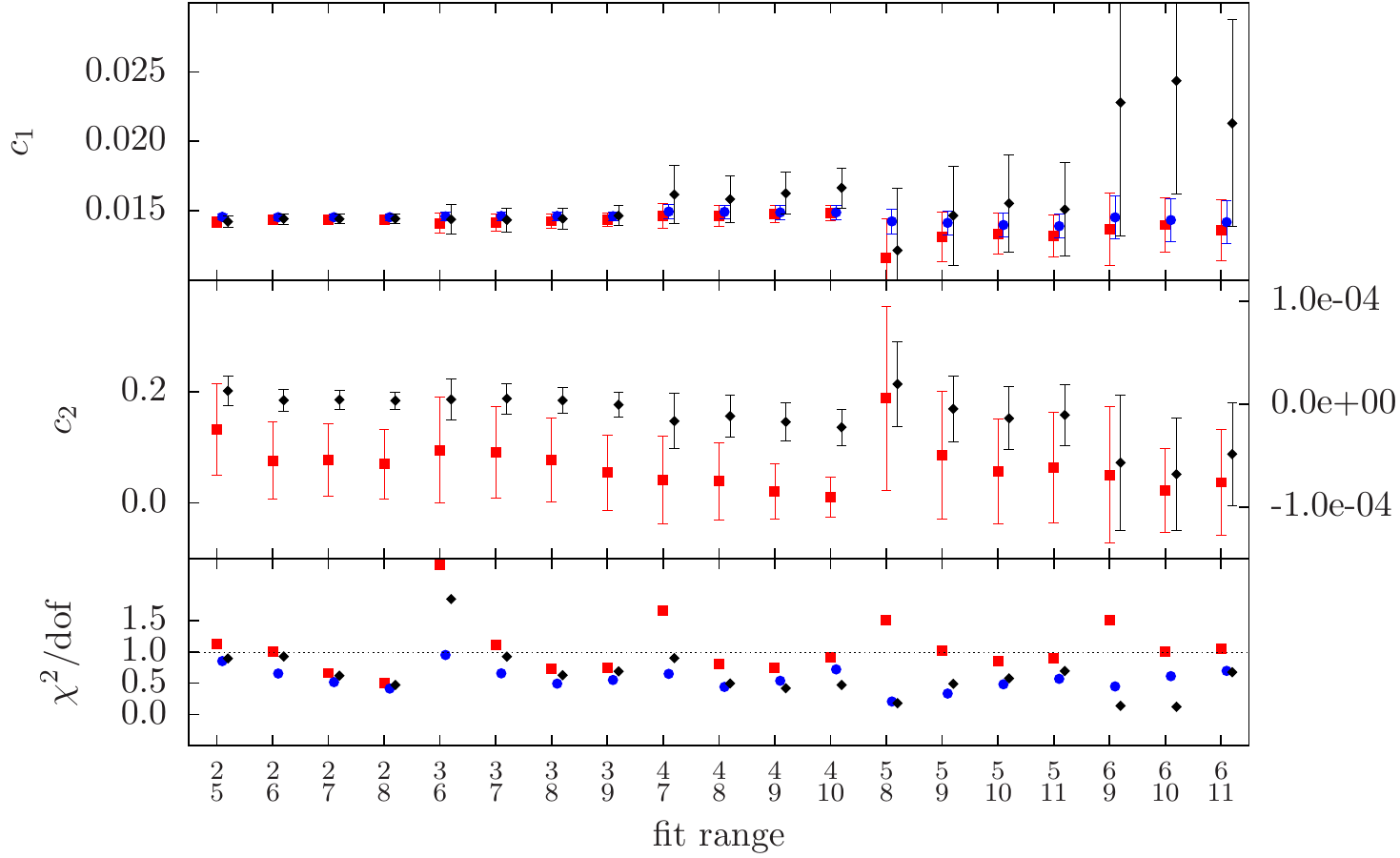}
  \includegraphics[width=0.47\textwidth]{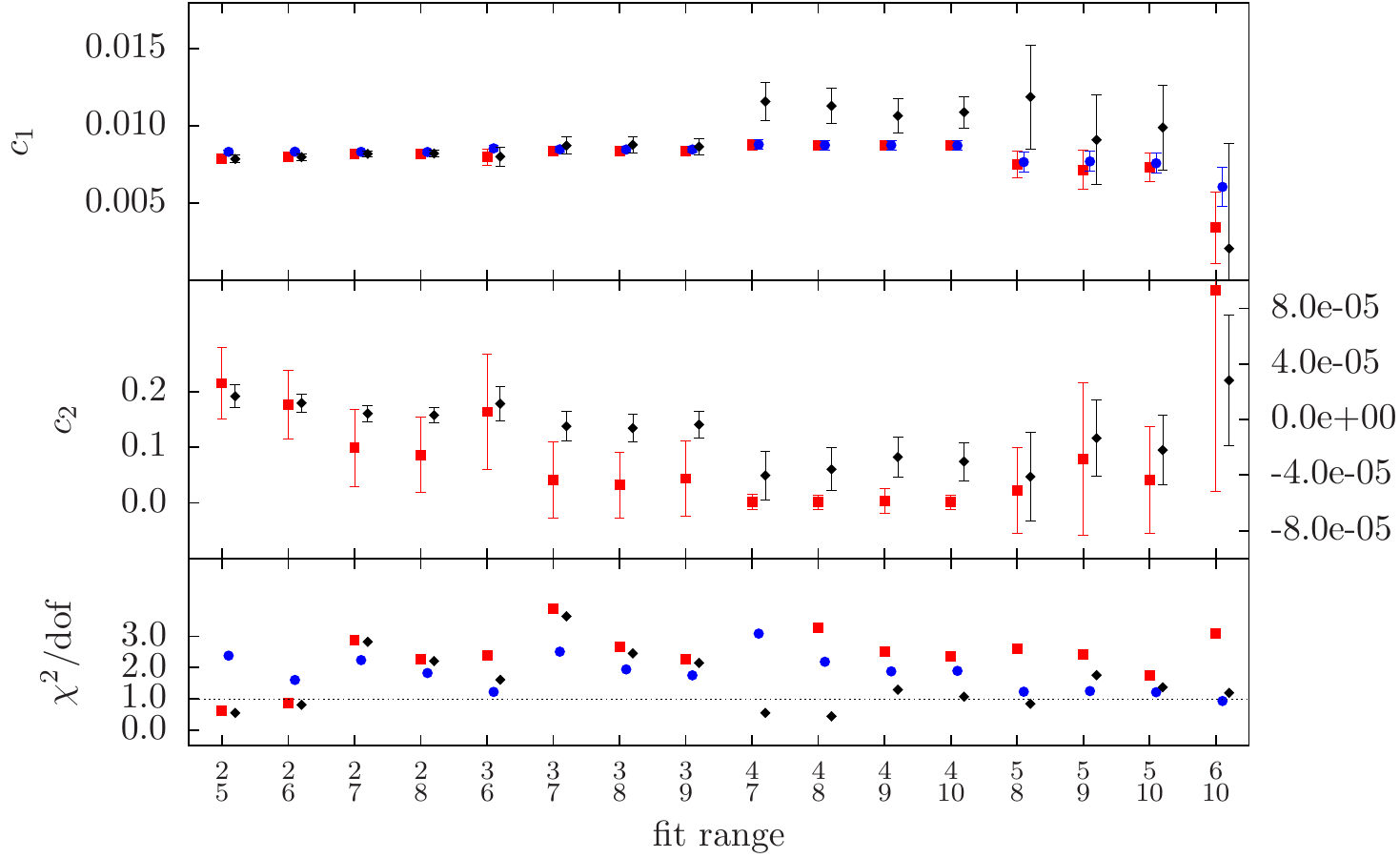}
  \includegraphics[width=0.47\textwidth]{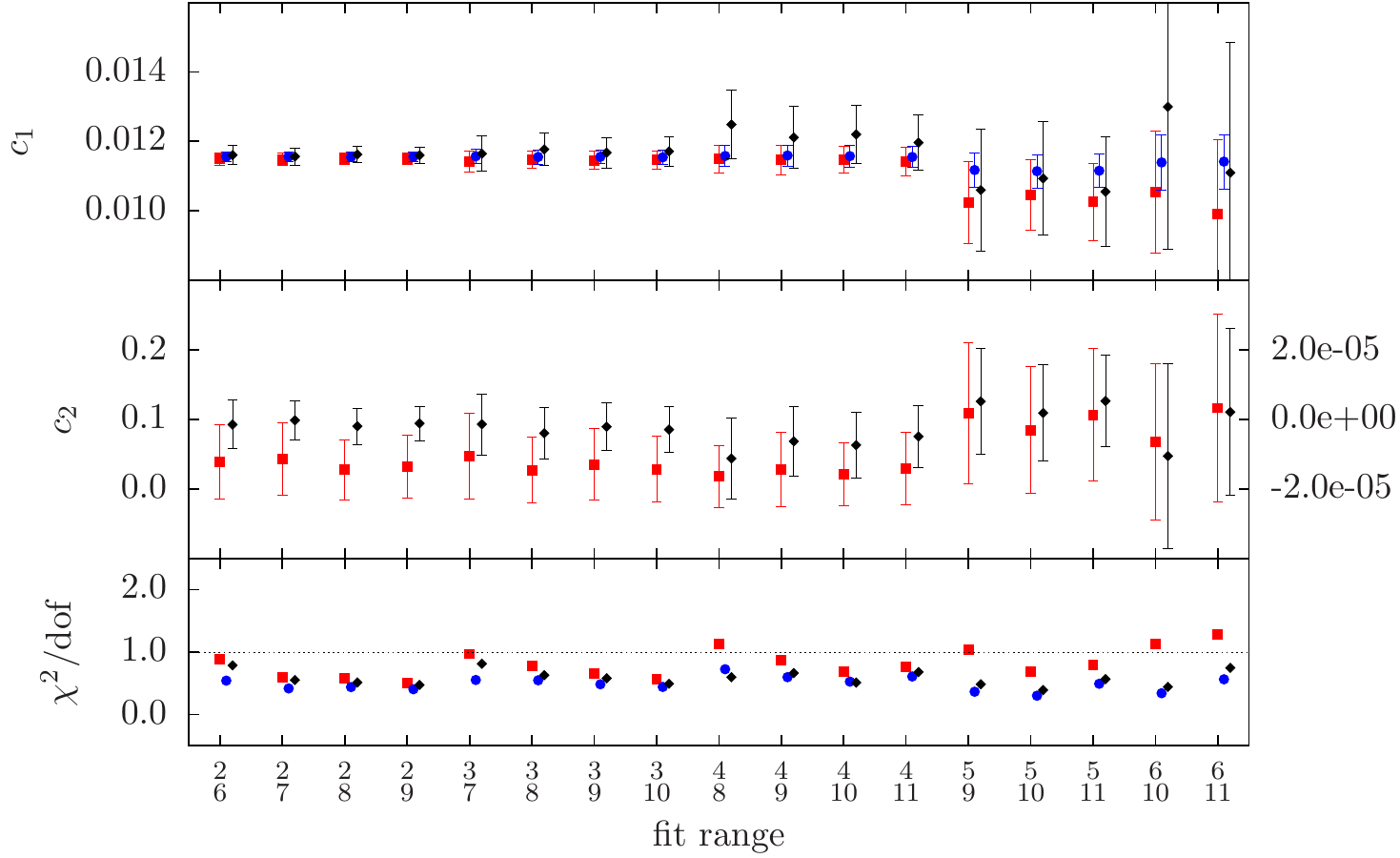}
  \caption{Samples of fit results for all four considered ratios for the unitary action: $R^{\Delta}_{\pi\,N}$ (top left), $R^{\Sigma^*}_{\pi\,\Lambda}$ (top right),
    $R^{\Sigma^*}_{\pi\,\Sigma}$ (bottom left) and $R^{\Xi^*}_{\pi\,\Xi}$ (bottom right). The notation is the same as that of Fig.~\protect\ref{fig:ratio_deltapp2piN_comparison}.}
  \label{fig:cmratio_unitary_fitresults_comparison}
\end{figure}

In Fig.~\ref{fig:cmratio_unitary_fitresults_comparison} we compile sample fit results for the unitary action for all four transition amplitudes. We find that the data cannot
constrain more than 3 fit parameters. Including $c_2$ already leads to high correlation among the fit parameters and large fluctuations of their values in the bootstrap sampling.
In addition, for the two transitions $\Delta \to \pi\,N$ and $\Sigma^* \to \pi\,\Lambda$ we can find a range of fit intervals, where we obtain $\chi^2/\mathrm{dof} \sim 1$ and
stable results for the parameter $c_1$ across the different fit ans\"atze and fit ranges.
For $\Sigma^* \to \pi\,\Sigma$
we observe larger changes in the parameters as we vary the fit ranges and very few fit ranges with a $\chisqrPerDof \approx 1$.

The coupling is  determined using the value of $c_1$ extracted from the fits
via  Eq.~\refeq{eq:width_formula} 
\begin{equation} 
 g^{B^*}_{M B} = c_1 \,\frac{\sqrt{N_{B^*}\,N_{M B}}}{V C_\mathrm{CG}}\,\frac{2 m_B}{|\kvec|}\,\left( \frac{1}{3}\,\frac{E_B(\kvec^2) + m_B}{m_B} \right)^{-1/2},
 \,\,\, \Gamma^{B^*}_{M\,B} = 2\pi\,\left[ \frac{2\,c_1^2}{2 s_{B^*}+1}\right]\,\rho(E_{MB})\,,
  \label{eq:width_formula}
\end{equation}
where $s_{B^*} = 3/2$ denotes the spin quantum number of the decuplet baryon.

\vspace*{-0.3cm}

\section{Conclusions and Outlook}

\vspace*{-0.3cm}

We compile our results for the coupling and width for all investigated transitions in Tables~\ref{tab:coupling_results} and \ref{tab:width_results}. 
In Table~\ref{tab:coupling_results} we compare the results for the coupling constant for the unitary  and 
the hybrid actions with the coupling at leading order in effective field theory (LO EFT) using as input the experimental values for the width and masses of 
of pion, decuplet and octet baryons~\cite{Beringer:1900zz}. In our calculation with both the hybrid and the unitary domain wall actions, the coupling
constants are close to the value obtained using LO EFT. This is remarkable given the size of the pion mass especially for the hybrid action, which is $350\mev$
and the non-degeneracy of energy levels.

\begin{table}
  \centering
\begin{tabular}{l|cc|c}
process & unitary & hybrid & PDG \\
	\hline
$\Delta^{++} \leftrightarrow \pi^+\,N^+$      & $ 23.7\,( 0.7)\,( 1.1)$ & $ 26.7\,( 0.6)\,( 1.4)$ & $29.4\,(0.3)$ \\
$\Sigma^{*+} \leftrightarrow \pi^+\,\Lambda$  & $ 18.5\,( 0.3)\,( 0.5)$ & $ 23.2\,( 0.6)\,( 0.8)$ & $20.4\,(0.3)$ \\
$\Sigma^{*+} \leftrightarrow \pi^+\,\Sigma^0$ & $ 16.1\,( 0.3)\,( 1.9)$ & $ 19.0\,( 0.7)\,( 2.9)$ & $17.3\,(1.1)$ \\
$\Xi^{*-} \leftrightarrow \pi^-\,\Xi^0$       & $ 21.0\,( 0.3)\,( 0.3)$ & $ 25.6\,( 0.6)\,( 4.3)$ & $19.4\,(1.9)$ \\
\hline
\hline
\end{tabular}

\caption{Results for the coupling $g^{B^*}_{M\,B}$. For each transition (first column) we show our result for the coupling with the unitary (second column) and the hybrid action (third column).
The fourth column gives the coupling based on experimental values for the baryon and meson mass and the decay width \cite{Beringer:1900zz} using leading order effective field theory.
The uncertainties are statistical and systematic.}
  \label{tab:coupling_results}
\end{table}
 \begin{table}
  \centering
  \begin{tabular}{l|cc|c}
process & unitary & hybrid & PDG \\
\hline
$\Delta^{++} \leftrightarrow \pi^+\,N^+$      & $ 119.4\,( 7.9)\,( 4.5)$ & $238.5\,( 12.2)\,( 16.2)$ & $118\,(2)$ \\
$\Sigma^{*+} \leftrightarrow \pi^+\,\Lambda$  & $  54.5\,( 2.1)\,( 1.3)$ & $ 143.9\,( 7.4)\,(  6.1)$ & $31.3\,(8)$ \\
$\Sigma^{*+} \leftrightarrow \pi^+\,\Sigma^0$ & $  17.6\,( 0.8)\,( 2.1)$ & $  58.3\,( 3.4)\,(  6.8)$ & $4.2\,(5)$ \\
$\Xi^{*-} \leftrightarrow \pi^-\,\Xi^0$       & $  35.1\,( 1.1)\,( 0.4)$ & $ 126.0\,( 5.6)\,( 18.5)$ & $9.9\,(1.9)$ \\
\hline
\hline
\end{tabular}

  \caption{Results for the decay widths $\Gamma^{B^*}_{M\,B}$. We compare our result for the unitary (second column) and hybrid (third column)
    action with the experimental value \cite{Beringer:1900zz} (fourth column).
  }
  \label{tab:width_results}
\vspace*{-0.3cm}
\end{table}
For the decay widths, given in Table~\ref{tab:width_results}, we find very good agreement with the experimental value for the $\Delta \to \pi\,N$ transition when computed using  the unitary action.
This corroborates the expectation, that the degeneracy of energy levels is a vital condition for the applicability of the transfer matrix method.
For the transition $\Sigma^* \to \pi\,\Lambda$ this condition is less well fulfilled and the decay width is off by 50\%. 
Moreover, the relative momentum in the $\pi-B-$state is fixed in physical units to $|\qvec| = 2\pi/L$ by the lattice parameters, which amounts to
$\approx 270\mev$ for the unitary action and $\approx 360\mev$ for the hybrid action. For the transition $\Delta \to \pi\,N$ with the unitary action
this is close to the corresponding experimental value of $227\mev$. For the three remaining decays $\Sigma^* \to \pi\,\Lambda,\,\pi\,\Sigma$ and $\Xi^* \to \pi\,\Xi$
the experimental value is $205\mev,\,120\mev$ and $158\mev$, respectively, and thus at greater discrepancy with the lattice kinematics.

From this study we conclude that the transfer matrix method is  a viable approach for the calculation of couplings and widths for a single channel  transition 
provided the respective energy levels are almost degenerate and the lattice kinematics close to the physical ones. The next steps to consider towards
an improved implementation of the method, are the addition of the quark-disconnected diagrams for the meson-baryon two-point correlation function and the inclusion of moving frames.
The latter in particular will  give a handle on both the size of the energy gap between initial and final state and on the relative momentum of the meson-baryon state.

\noindent
{\bf Acknowledgments:} This work is partly supported by
the U.S. DOE under grants DE-SC0011090, ER41888, and
DE-AC02-05CH11231, and the RIKEN Foreign Postdoctoral Researcher Program.
The computing resources were provided by the National Energy Research Scientific
Computing Center supported by the Office
of Science of the DOE under Contract No. DE-AC02-05CH11231,
the J\"ulich Supercomputing Center,
and by the Cy-Tera machine at the Cyprus Institute supported in
part by the Cyprus Research Promotion Foundation  under contract
NEA Y$\Pi$O$\Delta$OMH/$\Sigma$TPATH/0308/31.
The multi-GPU domain wall inverter code~\cite{Strelchenko:2012aa} is based on the QUDA library~\cite{Clark:2009wm,Babich:2011np}
and its development has been supported by PRACE grants RI-211528 and FP7-261557.

\vspace*{-0.3cm}

\end{document}